\newif\iftechreport
\techreportfalse

\iftechreport
\documentclass[orivec,runningheads]{llncs-href}
\else
\documentclass[orivec]{llncs} 
\fi

\usepackage{latexsym,amsmath,amssymb,amstext,amsfonts
}
\usepackage{stmaryrd}
\usepackage{listings}
\usepackage{semantic}
\usepackage{graphicx}
\usepackage{xyling} 
\usepackage{wrapfig}
\usepackage{url}

\DeclareMathAlphabet{\mathsc}{OT1}{cmr}{m}{sc}

\newcommand{\Set}[1] 
           {\left\{#1\right\}}

\newcommand{\SetDef}[2] 
           {\left\{#1\mathrel{\mid}#2\right\}}

\lstdefinelanguage{WHILE}{
  keywords={output, mod, div, to, int, while, if, then, else, for,  do, skip, declassify},
}

\mathlig{<<}{\langle}
\mathlig{>>}{\rangle}

\mathlig{|=>}{\Downarrow} 

\mathlig{<==>}{\Longleftrightarrow}

\mathlig{-->¤}{\stackrel{\bullet}{\rightarrow}}

\mathlig{-->e}{\stackrel{\tau}{\rightarrow}} 
\mathlig{-->*}{\mathbin{{\stackrel{\tau}{\rightarrow}}\!\rule{0pt}{0.4em}^{*}}}

\mathlig{-->0}{\stackrel{0}{\rightarrow}}
\mathlig{-->1}{\stackrel{1}{\rightarrow}}
\mathlig{-->o}{\stackrel{o}{\rightarrow}}

\mathlig{~>}{\leadsto}

\mathlig{|->}{\mapsto}

\newcommand{\PC}{Transcript} 

\pdfpagesattr{/CropBox [92 82 533 788]} 

\iftechreport
\usepackage{hyperref}
\usepackage{epsfig}
\fi

\begin{document}

\iftechreport

\begin{titlepage}
\begin{center}

{\Large Technical Report no. 2009-13 }
\\[10ex]
\textbf{\LARGE Specification and Verification of Side Channel Declassification${}^1$}
\\[5ex]
\textbf{\Large
Josef Svenningsson
\qquad
David Sands
}
\\
{
  \vspace{\stretch{1}} 
  \includegraphics{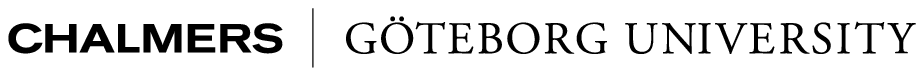} \\
  \vspace{5mm}
  \includegraphics{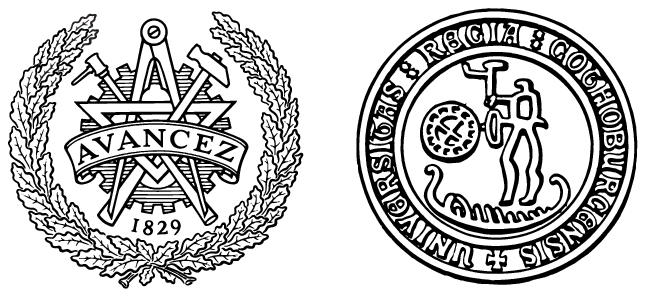} \\
  \vspace{12mm}
  Department of Computing Science and Engineering\\
  Chalmers University of Technology
  and Unversity of Gothenburg \\
  S-412 96 G\"{o}teborg, Sweden \\
  \vspace{8mm} G\"{o}teborg, December  2009 }
\vspace{40mm}
\end{center}

1. An abbreviated version of this article will appear in the proceedings
of \emph{Formal Aspects of Security and Trust} FAST 2009, to be
published in LNCS.

\end{titlepage}
\newpage
\thispagestyle{empty}
\mbox{}
\vspace{\stretch{1}}

\noindent
{\large
  \begin{tabular}{l}
    \includegraphics{Logos/ChalmGUmarke.eps} \\[3ex]
    Technical Report in Computing Science and Engineering at\\
    Chalmers University of Technology and University of Gothenburg
    \vspace{3ex} \\
    Technical Report no. 2009-13
\\    ISSN: 1650-3023 
 \\   \vspace{3ex} \\
    Department of Computing Science and Engineering\\
    Chalmers University of Technology
      and University of Gothenburg \\
    SE-412 96 G\"{o}teborg, Sweden
    \vspace{3ex} \\
    G\"{o}teborg, Sweden, 2009
  \end{tabular}
}
\newpage
\begin{samepage}
   \tableofcontents
\vspace{2ex}
   \begin{center}
     \Large \bf Abstract
   \end{center}\vspace{2ex}

  Side channel attacks have emerged as a serious threat to the
  security of both networked and embedded systems -- in particular through
  the implementations of cryptographic operations. Side channels can
  be difficult to model formally, but with careful coding and program
  transformation techniques it may be possible to verify security in
  the presence of specific side-channel attacks. But what if a program
  intentionally makes a tradeoff between security and efficiency and
  leaks some information through a side channel? In this paper we
  study such tradeoffs using ideas from recent research on
  declassification. We present a semantic model of security for
  programs which allow for declassification through side channels, and
  show how side-channel declassification can be verified using
  off-the-shelf software model checking tools. Finally, to make it
  simpler for verifiers to check that a program conforms to a
  particular side-channel declassification policy we introduce a
  further tradeoff between efficiency and verifiability: by writing programs
  in a particular ``manifest form'' security becomes considerably easier to 
  verify.

\end{samepage}
\newpage
\setcounter{footnote}{0}
\fi

\title{Specification and Verification of Side Channel Declassification
}

\author{Josef Svenningsson \and David Sands}

\institute{Department of Computer Science and Engineering,\\
           Chalmers University of Technology\\
           G\"oteborg, Sweden\\
	   \email{\{josefs,dave\}@chalmers.se}
}

\maketitle

\begin{abstract}


  Side channel attacks have emerged as a serious threat to the
  security of both networked and embedded systems -- in particular through
  the implementations of cryptographic operations. Side channels can
  be difficult to model formally, but with careful coding and program
  transformation techniques it may be possible to verify security in
  the presence of specific side-channel attacks. But what if a program
  intentionally makes a tradeoff between security and efficiency and
  leaks some information through a side channel? In this paper we
  study such tradeoffs using ideas from recent research on
  declassification. We present a semantic model of security for
  programs which allow for declassification through side channels, and
  show how side-channel declassification can be verified using
  off-the-shelf software model checking tools. Finally, to make it
  simpler for verifiers to check that a program conforms to a
  particular side-channel declassification policy we introduce a
  further tradeoff between efficiency and verifiability: by writing programs
  in a particular ``manifest form'' security becomes considerably easier to 
  verify.

\end{abstract}

\section{Introduction}
One of the pillars of computer security is confidentiality -- keeping
secrets secret. Much recent research in \emph{language based security} has
focused on how to ensure that information flows within programs do
not violate the intended confidentiality properties
\cite{Sabelfeld:Myers:JSAC}. 
One of the difficulties of tracking information flows is that
information may flow in various indirect ways.  Over 30 years ago,
Lampson \cite{Lampson:Confinement} coined the phrase \emph{covert
channel} to describe channels which were not intended for information
transmission at all.  At that time the concern was unintended
transmission of information between users on timeshared mainframe
computers. In much security research that followed, it was not
considered worth the effort to consider covert channels.  But with the
increased exposure of sensitive information to potential attackers,
and the ubiquitous use of cryptographic mechanisms, covert channels
have emerged as a serious threat to the security of modern systems --
both networked and embedded. The following key papers provide a view
of the modern side-channel threat landscape:
\begin{itemize}
\item Kocher \cite{Kocher:Timing} showed that by taking timing
measurements of RSA cryptographic operations one could discover secret
keys. Later \cite{Kocher+:Differential} it was shown that one could do
the same by measuring power consumption.
\protect
\item Based on Kocher's ideas numerous smart card implementations of
  cryptographic operations have shown to be breakable. See e.g.
  \cite{Messergers+:Power99}.

\item Brumley and Boneh \cite{Brumley:Boneh:Remote05} showed that
  timing attacks were not just relevant to smart cards and other
  physical cryptographic tokens, but could be effective across a
  network; they developed a remote timing attack on an SSL library
  commonly used in web servers.
\end{itemize}

What is striking about these methods is that the attacks are on
the \emph{implementations} and not features of the basic intended
functionality. Mathematically, cryptographic methods are adequately
secure, but useless if the functionally correct implementation has
timing or other side channels. 





\subsection{Simple Timing Channels}

Timing leaks often arise from the fact that computation involves
branching on the value of a secret. Different instructions are
executed in each branch, and these give rise to a timing leak or a
power leak (whereby a simple power analysis \cite{Mayer-Sommer:Smartly} can
reveal information about e.g. control flow paths). 

One approach is to ensure that both branches take the same time 
\cite{Agat:Timing}, or to eliminate branches altogether 
\cite{Molnar+:Program} -- an approach that is also well known from 
real-time systems where it is used to make worst case execution time easy to determine \cite{Puschner:Burns:Writing}.

\begin{wrapfigure}{l}{0.3\textwidth}
\begin{lstlisting}
r = 1;
i = m - 1;
while (i >= 0) {
  r = r * r;
  if (d[i] == 1) {
    r = r * x;
  }
  i = i - 1;
}
return r;
\end{lstlisting}
\caption{Modular exponentiation}
\label{modexp}
\end{wrapfigure}
Consider the pseudocode in Figure~\ref{modexp} representing a
na\"ive implementation of modular exponentiation, which we will use as
our running example throughout the paper.

The data that goes in to this function is typically secret. A common
scenario is that the variable \lstinline!x! is part of a secret which
is to be encrypted or decrypted and variable \lstinline!d! is the key
(viewed here as an array of bits). It is important that these remain
secret. (On the other hand, \lstinline!m!, the length of the key, is
usually considered public knowledge.)

However, as this function is currently written it is possible to
derive some or all of the information about the key using either a
timing or power attack. The length of the loop will always reveal the
size of the key -- and this is accepted. 
In the body of the loop there is a
conditional statement which is executed depending on whether the
current bit in the key is set or not. This means that each iteration
of the loop will take different amount of time depending on the value
of the key.  A timing attack measuring the time it takes to compute
the whole result can be used to learn the hamming weight of the key,
i.e. the number of 1's. With control over the key and repeated runs
this is sufficient to leak the key \cite{Kocher:Timing}. A power
analysis could in principle even leak the key in a single run.

\subsection{Timing and Declassification}
Often the run-time cost of securing an algorithm against timing attacks using a general purpose method is higher than what we are prepared to pay.\\
\begin{wrapfigure}{l}{45mm}
\begin{lstlisting}
z[0] = r*r mod N
z[1] = r*r*M mod N
r = z[d[i]]
\end{lstlisting}
\end{wrapfigure}
 For example, using a table-lookup instead of a branch
\cite{Coron:Resistance} the conditional can be replaced by the code to
the left.
This fixes the timing leak, but the algorithm becomes considerably
slower -- even after eliminating the common subexpression.  Another
even more costly approach is Agat's \emph{cross copying} idea, whereby
(roughly speaking) every branch on a secret value 
\lstinline!if h then A else B!  
is transformed into 
\lstinline!if h then A;[B] else [A];B!
where \lstinline![A]! is a ghost copy of \lstinline!A! which takes
the same time to compute but otherwise has no
effect. There are optimisations of this approach using unification
\cite{Kopf:Mantel:Unification07}, or by making the padding
probabilistic \cite{DiPierro+:2008}, but efficiency wise the
improvements offered by those techniques are probably not sufficient
in this context.

\begin{wrapfigure}[11]{l}{0.4\textwidth}
\begin{lstlisting}
r = 1;
i = m - 1;
k = 0;
while(i >= 0) {
  r = r * (k ? x : r);
  k = k ^ d[i];
  i = i - (k ? 0 : 1);
}
\end{lstlisting}
\caption{\small Protected exponentiation}
\label{frenchmodexp}
\end{wrapfigure}
A potential solution to this tension between security and efficiency
is to make a tradeoff between the two. For this reason it is not
uncommon for algorithms to have \emph{some} side channel leakage.  An
example of this is the following variation on modular exponentiation,
adapted from \cite{Chevallier+:Low-cost}, which is intended to 
provide some (unspecified) degree of protection against simple
power analysis attacks (but still leaks the hamming weight of the key).
\vspace{1ex}

\paragraph{Research Goals and Approach}
Our research goal is to determine how to express this
tradeoff. There are three key issues to explore:
\begin{itemize}
\item Security Policies: how should we specify 
side-channel declassification?
\item Security Mechanisms: how to derive programs which achieve the tradeoff?
\item Security Assurance: how can we  show that programs satisfy a given policy,  with a rigorous specification and formal verification?
\end{itemize}

This paper deals primarily with the first and the third point. 

The first step -- a prerequisite to a rigorous specification -- is to
specify our attacker model.  A model sets the boundaries of our
investigation (and as always with covert channels, there are certainly
attacks which fall outside). We choose (as discussed in
Section~\ref{sec:baseline}) the \emph{program counter security model}
\cite{Molnar+:Program}. This model captures attackers performing
simple power and timing analysis.

To specify a security policy 
we turn to work on declassification. 
The concept of declassification has been developed specifically to
allow the programmer to specify 
what, where, or when a piece of information is allowed
to leak (and by whom). A simple example is a program which requires a password based login. For this program to work it must declassify (intentionally leak) the value of the comparison between the actual and the user supplied password strings.

Declassification has been a recent hot topic in information flow
security (see \cite{Sabelfeld:Sands:Declassification} for an
overview). 
The standard techniques for declassification seem largely applicable to our problem, but there are some differences. The reason being that (in the context of cryptographic algorithms
in particular) we may be interested in the distinction between
declassifying some data directly (something which has potentially zero
cost to the attacker), and declassifying the data but only through a
\emph{side channel} -- the latter is what we call \emph{side channel
  declassification}.

We will adapt existing declassification concepts to specify
\emph{what} information we are willing to leak through timing channels
(Section~\ref{sec:extensional}) . More specifically, we use small
programs as a specification of \emph{what} information is leaked. This
follows the style of \emph{delimited release}
\cite{Sabelfeld:Myers:ISSS03}.
As an example, we might want to specify
that a program does not leak more that the hamming weight of the
key. This can be achieved by using the program fragment in Figure~\ref{hamming}
as a specification: it explicitly computes the hamming weight of the key.

\begin{wrapfigure}{l}{0.4\textwidth}
\begin{lstlisting}
h = 0;
i = m - 1;
while (i >= 0) {
  if (d[i] == 1) {
    h = h + 1;
  }
  i = i - 1;
}
\end{lstlisting}
\caption{Hamming weight computation}
\label{hamming}
\end{wrapfigure}
The formal definition of side-channel declassification
(Section~\ref{sec:extensional}) is that if the attacker knows the
information leaked by the declassifier then nothing more is learned by
running the program.

We then turn to the question of verification. We investigate the use
of off-shelf automatic program verification tools 
to verify side-channel declassification
policies.  The first step is to reify the side channel by transforming
the program to represent the side-channel as part of the program state
(Section~\ref{sec:reify}).  This reduces the specification of
side-channel declassification to an extensional program property.

The next step is to observe that in many common cases we can simplify
the side-channel instrumentation. This simplification (described in
Section~\ref{sec:simplification}) does not need to be semantics
preserving -- it simply needs to preserve the side-channel
declassification condition.



As we aim to use automatic off-the-shelf model checkers we need one
final transformation to make our programs amenable to verification. We
use \emph{self composition} to reduce the verification problem to a
safety property of a transformed
program. Section~\ref{sec:selfcomposition} describes the approach and
experiments with software model checkers.

For various reasons the side-channel declassification property of
algorithms can still be hard to verify. The last part of this work
(Section~\ref{sec:manifest}) introduces a tradeoff which makes
verification much simpler. The idea is to write programs in what we
call \emph{manifest form}. In manifest form the program is written in
two parts: a declassifier first computes \emph{what} is to be
released, and then using this information a side-channel secure
program computes the rest. 

The verification problem amounts to showing that the second part of
the program is indeed side-channel secure (this can be rather
straightforward due to the strength of the side-channel security
condition),  and that the declassifier
satisfies the property that it does not leak more through its side
channel than it leaks directly. We call these \emph{manifest
declassifiers}. Since declassifiers are much simpler (and quite likely
useful in many different algorithmic contexts) verification of
manifest declassifiers is relatively simple. We show how this
technique can overcome the verification limitations of certain
verification tools.

\iftechreport
\else
An extended version of this article containing material left out for reasons
of space constraints is available as technical report no. 2009:13 from Department of Computing Science and Engineering at 
and from \url{arxiv.org/...}.
\url{www.cse.chalmers.se/~josefs}.
\fi

\section{Preliminaries}

In this section we present the language we are going to use and set up the
basic machinery in order to define our notion of security.

Since we target cryptographic algorithms we will be using a small
while language with arrays. It's syntax is defined below:
\[
\begin{array}{lcl}
C \in \text{Command} & ::= & x = e \; | \; x[y] = e \; | \; C_1 ; C_2 
\; | \; \text{if} \; e \; \text{then} \; C_1 \; \text{else} \; C_2 \; | \; \text{while} \; e \; C \; | \; \text{skip}\\
e \in \text{Expression} & ::= & x \; | \; x[e] \; | \; n \; | \; e_1 \; \text{op} \; e_2 \; | \; x \; ? \; y \; : \; z\\
op \in \text{Operators} & := & + \; | \; * \; | \; - \; | \; \text{\textasciicircum} \; | \; \text{mod} \; | \; \dots
\end{array}
\]

The commands of the program should not require much explanation as
they are standard for a small while language. 

One particular form of expression that we have chosen to include that
may not look very standard (for a toy language) is the ternary
operator borrowed from the language C. It's a conditional expression
that can choose between the value of two different register based on
the value of a third register. We have restricted it to only operate
on registers since allowing it to choose between evaluating two
general expressions may give rise to side channels. This kind of
operation can typically be implemented to take a constant amount of
time so that it doesn't exhibit a side channel by using conditional
assignment that is available in e.g. x86 machine code.


The semantics of programs is completely standard. We defer the definition until the next section where an operational semantics is given together with some additional instrumentation. 


\section{Baseline Security Model}
\label{sec:baseline}

In this section we present the semantic security model which we use to
model the attacker and to define the baseline notion of
declassification-free security. For a good balance between simplicity and 
strength we adopt an existing approach: the \emph{program counter security model} \cite{Molnar+:Program}. This attacker model
is strong enough to analyze \emph{simple power analysis} attacks
\cite{Kocher+:Differential} -- where the attacker is assumed to be
able to make detailed correlations between the power profile of a
single run with the instructions executed during that run.

The idea of the \emph{program counter security model} is to assume the
attacker can observe a \emph{transcript} consisting of the sequence of
program counter positions. This is slightly stronger than an attacker
who could perfectly deduce the sequence of instructions executed from
a (known) program given a power consumption profile of an execution.
It does, however, assume that the power consumption of a particular
operation does not depend on the data it manipulates. In particular it
does not model differential power analysis.

Suppose a program operates on a state which can be partitioned into a
\emph{low} (public) part, and a \emph{high} (secret) part. A program is
said to be \PC-secure if given any two states whose low parts
are equal, running the program on these respective states yields equal
transcripts and final states which also agree on their low parts.\footnote{It would be natural to assume that attackers have only 
polynomially bounded computing power in the size of the high part of the 
state. For the purposes of this paper our stronger definition will suffice.}

To specialise this definition to our language we note that it is
sufficient for the attacker to observe the sequence of branch
decisions in a given run in order to be able to deduce the sequence of
instructions that were executed. To this end, 
in Figure~\ref{fig:sidechannelsem} we give an instrumented semantics
for our language which makes this model of side channels concrete. 
Apart from the instrumentation (in the form of labels on the transitions) this is a completely standard small-step operational semantics. 
The transition labels, $o$,  are either a silent step ($\tau$), a \(0\) or a \(1\).  
A zero or one is used to record which branch was taken in an \lstinline!if! or \lstinline!while!
 statement. 
\begin{figure}
  \begin{gather*}<< n , S >> \Downarrow n
\qquad
<< x , S >> \Downarrow S(x)
\qquad
\inference{<< e, S >> \Downarrow v}
	  {<< x[e], S >> \Downarrow S(x)(v)}
\\ 
\inference{<< e_1 , S >> \Downarrow v_1 & << e_2 , S >> \Downarrow v_2}
	  {<< e_1 \; \text{op} \;  e_2 ,S >> \Downarrow v_1 \; \overline{\text{op}} \; v_2}
\qquad
\inference{S(x) \neq 0}{<< x ? y : z , S >> \Downarrow S(y)}
\qquad
\inference{S(x) = 0}{<< x ? y : z , S >> \Downarrow S(z)}
\\[2ex]
\inference{<< e , S >> \Downarrow v}
	  {<< x = e , S >> -->e << \text{skip} , S[x |-> v]>>}
\qquad
\inference{<< e , S >> \Downarrow v}
	  {<< x[y] = e , S >> -->e << \text{skip} , S[x |-> x[ S(y) |-> v ]] >> }
\\[1ex]
{ << \text{skip} ; C , S >> -->e << C , S >> }
\qquad 
\inference{<< C_1 , S >> -->o << C'_1 , S' >> }
	   { << C_1 ; C_2 , S >> -->o << C'_1 ; C_2 , S' >> }
\\[1ex]
\inference{<< e , S >> \Downarrow v & v \neq 0}
	  {<< \text{if} \; e \; C_1 \; C_2 , S >> -->1 << C_1 , S >>}
\qquad
\inference{<< e , S >> \Downarrow 0}
	  {<< \text{if} \; e \; C_1 \; C_2 , S >> -->0 << C_2 , S >>}
\\[1ex]
\inference{<< e , S >> \Downarrow v & v \neq 0}
	  {<< \text{while} \; e \; C ,S >> -->1 << C ; \text{while} \; e \; C , S >>}
\qquad 
\inference{<< e , S >> \Downarrow 0}
	  {<< \text{while} \; e \; C ,S >> -->0 << \text{skip} , S >>}
\end{gather*}
\label{fig:sidechannelsem}
\caption{Instrumented Semantics}
\end{figure}

\begin{definition}[Transcript]
  Let $d_1, d_2,\ldots$ range over $\{ 0, 1 \}$.  We say that a configuration
  $<< C , S >>$ has a transcript $d_1,\ldots,d_n$ if there exist configurations
$<<C_i,S_i>>$, $i \in [1,n]$ such that 
  \[
  << C , S >> -->* \stackrel{d_1}{\rightarrow} 
  << C_1 , S_1 >> -->* \stackrel{d_2}{\rightarrow} \cdots 
  -->* \stackrel{d_n}{\rightarrow} << C_n , S_n >>   -->* << \verb!skip!, S'>>
  \] 
for some $S'$.

In the above case we will write $|[ C |] S = S'$ 
(when we only care about the final state) and  
$|[ C |]^T S = (S',t)$  where $t = d_1,\ldots,d_n$ (when we are interested in the state and the transcript).  
\end{definition}
For the purpose of this paper (and the kinds of algorithms in which we
are interested in this context) we will implicitly treat $|[ C |]$ and
$|[ C |]^T$ as functions rather than partial functions, thus ignoring
programs which do not always terminate.

Now we can formally define the baseline security definition, which following \cite{Molnar+:Program} we call \PC-security:
\begin{definition}[\PC-Security]
Assume a partition of program variables into \emph{low} and \emph{high}. We write $R =_L S$ if program states $R$ and $S$ differ on at most their high variables. We extend this to state-transcript pairs by $(R,t_1) =_L (S,t_2) \iff R =_L S \And t_1 = t_2$ reflecting the fact that a transcript is considered attacker observable (low).  

A program $C$ is \emph{\PC-secure} if for all $R$, $S$, if $R =_L S$ then 
$|[ C |]^T R =_L  |[ C |]^T S$.
\end{definition}






Note that \PC-security, as we have defined it, is a very strong
condition and also very simple to check. A sufficient condition for
\PC-security is that the program in question (i) does not assign
values computed using high variables to low variables, and (ii) does
not contain any loops or branches on expressions containing high
variables. The main contribution of \cite{Molnar+:Program} is a suite
of methods for transforming programs into this form. Unfortunately the
transformation can be too costly in general, but that method is nicely
complemented by use of declassification.
 
\section{Side Channel Declassification}
\label{sec:extensional}

To weaken the baseline definition of security we adopt one of the
simplest mechanisms to specify \emph{what} information may be leaked
about a secret: \emph{delimited release}
\cite{Sabelfeld:Myers:ISSS03}. The original definition of delimited
release specified declassification by placing declassify labels on
various expressions occurring in a program. The idea is that the
attacker is permitted to learn about (at most) the values of those
expressions in the initial state, but nothing more about the high part
of the state.

We will reinterpret delimited release using a simple program rather
than a set of expressions. The idea will be to specify a (hopefully
small and simple) program $D$ which leaks information from high
variables to low ones. 
 A program is \PC-secure modulo
declassifier $D$ if it leaks no more than $D$, and this leak 
occurs \emph{through the side channel}. 
\begin{definition}[Side Channel Declassification]
Let $D$ be a program which writes to variables distinct from all variables 
occurring in $C$. 
We define  $C$ to be \emph{\PC-secure modulo $D$} if for all $R$ and $S$ such that $R =_L S$
we have 
\[
  |[ C |] R =_L |[ C |] S \And (|[ D |] R = |[ D |] S =>  |[ C |]^T R =_L |[ C |]^T S).
\]
\end{definition}
The condition on the variables written by $D$ is purely for convenience, but
is without loss of generality. 
The first clause of the definition says that the only information leak can be through the side channel. 
The second clause says that the leak is no more than what is directly leaked by $D$. It is perhaps helpful to consider this clause in contrapositive form:
\(
|[ C |]^T R \neq_L |[ C |]^T S  => |[ D |] R \neq |[ D |] S.
\) 
This means that if there is an observable difference in the transcripts of two runs then that difference is manifest in the corresponding runs of the declassifier.
Note that if we had omitted the condition  $|[ C |] R =_L |[ C |] S$
then we would have the weaker property that $C$ would be allowed
to leak either through the store or through the side channel -- but we
wouldn't know which. From an attackers point of view it might take
quite a bit more effort to attack a program if it only leaks though
the side channel so it seems useful to make this distinction.
Clearly there are other variations possible
involving multiple declassifiers each leaking through a particular
subset of observation channels.




\section{Reifying the side channel}
\label{sec:reify}
In the previous sections we have a definition of security that enables
us to formally establish the security of programs with respect to side
channel declassification. We now turn to the problem of verifying that
particular programs fulfil the security condition. In order to avoid
having to develop our own verification method we have chosen to use
off-the-shelf software verification tools.

Software verification tools work with the standard semantics of
programs. But recall that our security condition uses an instrumented
semantics which involves a simple abstraction of side channels.  In
order to make it possible to use off-the-shelf tools for our security
condition we must reify the transcript so that it becomes an explicit
value in the program which the tools can reason about. It is easy to
see how to do this: we add a list-valued variable $t$ to the program,
and transform, inductively, each conditional
 \lstset{backgroundcolor=\color{white}}
\lstinline!if e then C else C'! into 
\lstinline!if e then t = t++"1"; C else t = t++"0"; C'!
and each while loop \lstinline!while e do C! into
\begin{center}
  \lstinline!(while e do t = t++"1"; C); t= t++"0"!
\end{center} 
 \lstset{backgroundcolor=\color[rgb]{1,0.98,.98}}
and inductively transform the subexpressions \lstinline!C! and \lstinline!C'!.

\subsection{Simplifying the instrumentation}
\label{sec:simplification}
Reifying the transcript from the instrumented semantics in this way
will create a dynamic data structure (a list) which is not bounded in size
in general. Such data structures make programs more difficult to reason about,
especially if we want some form of automation in the verification
process. Luckily, there are several circumstances which help us side
step this problem. Concretely we use two facts to simplify the
reification of the side channel.

The first simplification we use depends on the fact that we do not
have to preserve the transcript itself -- it is sufficient that it
yields the same low-equivalence on programs.  Suppose that \(P^T\) is
the reified variant of the program \(P\) and that the reification is
through the addition of some low variables. In order to use \(P^T\) for
verification of side-channel security properties it is sufficient for
it to satisfy the following property:
\[
\forall R, S . |[ P |]^T R =_L |[ P |]^T S  <=> |[ P^T |] R =_L |[ P^T |] S
\]
We call such a $P^T$ an \emph{adequate} reification of $P$.

%
\begin{wrapfigure}[13]{rx}{0.4\textwidth}
\begin{lstlisting}[numbers=right]
r = 1;
i = m - 1;
k = 0; t = 0;
while(i >= 0) {
  t = t + 1;
  r = r * (k ? x : r);
  k = k xor d[i];
  i = i - (k ? 0 : 1);
}
\end{lstlisting}
\caption{\small Instrumented modular exponentiation}\
\label{instrumentedmodexp}
\end{wrapfigure}
The second simplification that we can perform in the construction of a
reified program is that we are specifically targeting cryptographic
algorithms. A common structure among the ones we have tried to verify
is that the while loops contain
straight line code (but potentially conditional expressions). 
If it is the case that 
\lstinline!while! loops don't contain any nested branching or looping
constructs then we can avoid introducing a dynamic data structure to
model the transcript. Let us refer to such programs as
\emph{unnested}. 
For unnested programs it is simply enough to use one fresh low variable
for each occurrence of a branch or loop. Thus the reification
transformation for unnested programs is defined by applying the two
transformation rules below to each of the loops and branches respectively:
\\[2ex]
\begin{tabular}{rcll}
  \lstinline!while e C! & $~>$ & \lstinline!v = 0; while e (v = v + 1; C)! &\quad ($v$ fresh)
\\[1ex]
  \lstinline!if e then C else C'! & $~>$ & \lstinline!if e then v = 1; C else v = 0; C'! & \quad ($v$ fresh)
\end{tabular}
\\[2ex]


The program in Figure~\ref{instrumentedmodexp} is an instrumented version of the program in
Figure~\ref{frenchmodexp}. The only change is the new (low) variable
\lstinline!t! which keeps track of the number of iterations in the
\lstinline!while! loop.

\section{Self Composition}
\label{sec:selfcomposition}
Standard automatic software model checking tools cannot reason about
multiple runs of a program. They deal exclusively with safety
properties which involves reasoning about a single run.  As is
well-known, noninterference properties (like side-channel
declassification) are not safety properties -- they are defined as
properties of pairs of computations rather than individual ones.
However, a recent technique has emerged to reduce noninterference
properties to safety properties for the purpose of verification. The
idea appeared in \cite{Darvas+:WITS03}, and was explored extensively
in \cite{Barthe+:CSFW04} where the idea was dubbed \emph{self
composition}.  Suppose $C$ is the program for which we want to verify
noninterference. Let $\theta$ be a bijective renaming function to a
disjoint set of variables from those used in $C$. Let $C_\theta$
denote a variable renamed copy of $C$. Then the standard
noninterference property of $C$ can be expressed as a safety property
of $C;C_\theta$ viz. the Hoare triple
\[
\{ \forall v \in \mathit{Low}. v = \theta(v) \} C;C_\theta \{ \forall v \in \mathit{Low}. v = \theta v \}
\]
To extend this to deal with side-channel declassification, let us
suppose that $C^T$ is an adequate reification of $C$.  Then we can
verify \PC-security modulo $D$ by the Hoare triple above (non
side-channel security) in conjunction with:
\[
\{ \forall v \in \mathit{Low}. v = \theta(v) \} D;D_\theta;C^T;C^T_\theta \{ (\forall x \in W.x = \theta(x)) => \forall y \in \mathit{Low}.y = \theta(y) \}
\]
where $W$ denotes the variables written by $D$. Here we take advantage
of the assumption that the variables written by $D$ are disjoint from
those used in $C^T$. This enables us to get away with a single
renaming.  Note that since $D$ is a program and not an expression we
cannot simply use it in the precondition of the Hoare triple
(c.f. \cite{Barthe+:CSFW04,Terauchi:Aiken:SAS05}).

\subsection{Experiments using Self Composition}

As Terauchi and Aiken discovered when they used self composition, it
often resulted in verification problems that were too hard for the
model checkers to handle \cite{Terauchi:Aiken:SAS05}. As a result of
this they developed a series of techniques for 
making the result of self composition
easier to verify. The main technique is the observation that the low part of
the two initial states must be equal and hence any computation that
depends only on the low part can safely be shared between the two
copies of the program. This was reported to help verifying a number of
programs. We employ the same technique in our experiments.

We have used the model checkers Blast\cite{Blast:2003} and
Dagger\cite{Dagger:2006} and applied them to self composed version of
the cryptographic algorithms. In particular we have tried to verify
the instrumented modular exponentiation algorithm in
Figure~\ref{instrumentedmodexp} secure modulo the hamming weight of
the key (Figure~\ref{hamming}). 
\iftechreport
Appendix~\ref{sec:code} presents the code given to the model checkers. 
\fi
We have also tried all the algorithms
proposed in \cite{Chevallier+:Low-cost} since they all exhibit some
form of side-channel leak and therefore have to be shown to be secure
relative that leak. None of the model checkers were powerful enough to
automatically verify the programs secure.

The main reason these tools fail seems to be that they do not reason
about the contents of arrays. Being able to reason about arrays is
crucial for our running example, as it involves computing the hamming
weight of an array.

Another problem comes from the fact that the programs we wish to prove
secure may be very different from its declassifier. Relating two
different programs with each other is a very difficult task and not
something that current software model checkers are designed to do.

By helping the model checkers with some manual intervention it is
possible to verify the programs secure. Blast has a feature which
allows the user to supply their own invariants. Given the correct
invariants it will succeed with the verification. However, these
predicates are not checked for correctness and coming up with them can
be a highly non-trivial task. We have therefore developed another
method for verification which we will explore in the next section.

\section{Manifest form}
\label{sec:manifest}

In this section we introduce a new way to structure programs to make
verification considerably easier: \emph{Manifest Form}. In manifest
form the program is written in two parts: a declassifier first
computes \emph{what} is to be released, and then using this
information a \PC-secure program computes the rest. Manifest form
represents a tradeoff: writing a program in manifest form may make it
less efficient. The idea is that the program makes the
declassification explicit in its structure (this is similar to the
specification of \emph{relaxed noninterference}
\cite{Li:Zdancewic:POPL05}).  
But for this to be truly explicit declassification 
the declassifier itself should not leak through its side channel
 -- or more precisely, the declassifier should not leak
more through its side channel than it does directly through the store.
\begin{definition}[Manifest Declassifier]
A program \(D\) is said to be a \emph{Manifest Declassifier} if
for all $R$ and $S$
\[
 |[ D |] S =_L |[ D |] R => |[ D |]^T S =_L |[ D |]^T R
\]
\end{definition}
As an example of a \emph{non} manifest declassifier, consider the
program to the left below which declassifies whether an array of
length \lstinline!m! contains all zeros.  Here the array length
\lstinline!m!, and \lstinline!i! and the declassified value
\lstinline!allz!, are low.  This is not manifest because the
transcript leaks more than the store: it reveals the position of the
first nonzero element. A manifest version of this declassifier is shown
on the right:
\begin{minipage}[t]{0.5\textwidth}
\begin{lstlisting}
i = m - 1; allz = 1;
while(allz and i >= 0) {
  allz = (d[i]? 0 : 1);
  i = i - 1;
}
i = 0
\end{lstlisting}
\end{minipage}
 ~ \vline ~~  
\begin{minipage}[t]{0.42\textwidth}
\begin{lstlisting}[numbers=right]
i = m - 1; allz = 1;
while(i >= 0) {
 allz *= (d[i]? 0 : 1);
 i = i - 1;
}
\end{lstlisting} 
\end{minipage}

\begin{definition}[Manifest Form] 
A program \(P\) is in \emph{Manifest Form} if \(P = D;Q\) where \(D\)
is a manifest declassifier and \(Q\) is transcript secure.
\end{definition}

The program in Figure~\ref{manifestmodexp} 
is written in manifest form but otherwise it represents the
same algorithm as the program in Figure~\ref{frenchmodexp}. The first
part of the program (lines 1--6) computes the hamming weight of the key,
\lstinline!d!, and this (using low variable \lstinline!hamming!)
is then used in the second part of the program
to determine the number of loop iterations.
\begin{figure}[htb]
\begin{minipage}[t]{0.5\textwidth}
\vspace{1ex}
\begin{lstlisting}[frame=topline,numbers=left]
hamming = 0; 
i = m - 1; 
while(i >= 0) { 
  hamming += (d[i] ? 1 : 0);
  i = i + 1; 
}
\end{lstlisting}
\end{minipage}
\qquad
\begin{minipage}[t]{0.4\textwidth}
\begin{lstlisting}[frame=bottomline,numbers=left,firstnumber=7]
r = 1; k = 0; 
i = m - 1; 
j = m - 1 + hamming; 
while(j >= 0) { 
  r = r * (k ? x : r); 
  k = k xor d[i]; 
  i = i - (k ? 0 : 1); 
  j = j - 1; 
}
\end{lstlisting}
\end{minipage}
\caption{Modular Exponentiation in Manifest Form}
\label{manifestmodexp}
\end{figure}


\iftechreport
Another example of a program in manifest form can be found in
Appendix~\ref{sec:manifestexample}.
\fi

\subsection{Manifest Security Theorem}
Armed with the definitions of sound manifest declassifiers we can
now state the theorem which is the key to the way we verify side-channel declassification.
\begin{theorem}
Given a program \(P = D;Q\) with \(D\) being a sound manifest
declassifier and \(Q\) is transcript secure then \(P\) is transcript
secure modulo \(D\)
\end{theorem}
This theorem helps us decompose and simplify the work of verifying
that a program in manifest form is secure. First, showing that \(Q\)
is transcript secure is straightforward as explained in
section~\ref{sec:baseline}. Verifying that \(D\) is a sound manifest
declassifier, which might seem like a daunting task given the
definition, is actually something that is within the reach of current
automatic tools for model checking. 
\iftechreport
We present the code we used to
verify the hamming weight computation in Appendix~\ref{sec:code}.
\fi

We apply the same techniques of reifying the side channel and self
composition to the problem of verifying sound manifest
declassifiers. When doing so we have been able to verify that our
implementation of the hamming weight computation in
Figure~\ref{hamming} is indeed a sound manifest declassifier and
thereby establishing the security of the modular exponentiation
algorithm in Figure~\ref{manifestmodexp}. We have had the same success\footnote{Using Blast version 2.5}
with all the algorithms presented in
\cite{Chevallier+:Low-cost}. 

\section{Related Work}
The literature on
programming language techniques for information flow
security is extensive. Sabelfeld and Myers survey
\cite{Sabelfeld:Myers:JSAC} although some seven years old remains the standard
reference in the field. It is notable that almost all of the work in
the area has ignored timing channels. However any automated security
checking that does not model timing will accept a program which leaks
information through timing, no matter how blatant the leak is.

Agat \cite{Agat:Timing} showed how a type system for secure information
flow could be extended to also transform out certain timing leaks by
padding the branches of appropriate conditionals.  K\"opf and Mantel
give some improvements to Agat's approach based on code unification
\cite{Kopf:Mantel:Unification07}. In a related line, Sabelfeld and
Sands considered timing channels arising from concurrency, and made
use of Agat's approach \cite{Sabelfeld:Sands:CSFW00}. Approximate and
probabilistic variants of these ideas have also emerged
\cite{DiPierro+:Tempus,DiPierro+:2008}.   The
problem with padding techniques in general is that they do not change
the fundamental structure of a leaky algorithm, but use the
``worst-case principle'' \cite{Agat:Sands:SSP01} to make all
computation paths equally slow.  For cryptographic algorithms this
approach is probably not acceptable from a performance perspective.

Hedin and Sands \cite{Hedin:Sands:Bytecode,Hedin:LIC} consider
applying Agat's approach in the context of Java bytecode. One notable
contribution is the use of a family of \emph{time models} which can
abstract timing behaviour at various levels of accuracy, for example
to model simple cache behaviour or instructions whose time depends on
runtime values (e.g. array allocation). The definitions and analysis
are parameterised over the time models.  The program counter security
model \cite{Molnar+:Program} can be seen as an instance of this
parameterised model.

More specific to the question of declassification and side channels,
as we mentioned above, \cite{DiPierro+:2008} estimates the capacity of
a side channel -- something which can be used to determine whether the
leak is acceptable -- and propose an approximate version of Agat's
padding technique.  Giacobazzi and Mastroeni
\cite{Giacobazzi:Mastroeni:Timed} recently extended the abstract
noninterference approach to characterising what information is leaked
to include simple timing channels. Their theoretical framework could
be used to extend the present work. In particular they conclude with a
theoretical condition which, in principle, could be used to verify
manifest declassifiers.  K\"opf and Basin's study of timing channels
in synchronous systems\cite{Kopf:Basin:Timing} is the most closely
related to the current paper. They study a Per model for expressing
declassification properties in a timed setting -- an abstract
counterpart to the more programmer-oriented delimited release approach
used here. They also study verification for deterministic systems by
the use of reachability in a product automaton -- somewhat analogous
to our use of self composition. Finally their examples include leaks
of hamming weight in a finite-field exponentiation circuit.

\section{Conclusions and Further Work}
Reusing theoretical concepts and practical verification tools we have
introduced a notion of side channel declassification and shown how
such properties can be verified by a combination of simple
transformations and application of off-the-shelf software model
checking tools. We have also introduced a new method to specify
side-channel declassification, \emph{manifest form}, a form which
makes the security property explicit in the program structure, and
makes verification simpler. We have applied these techniques to verify
the relative security of a number of cryptographic algorithms.  It
remains to investigate how to convert a given program into manifest
form. Ideas from \cite{Molnar+:Program,Li:Zdancewic:POPL05} may be
adaptable to obtain the best of both worlds: a program without the
overhead of manifest form, but satisfying the same side-channel
declassification property. 

\small
\bibliographystyle{alpha} 
\bibliography{bib}

\iftechreport

\newpage

\appendix

\section{Example Code Used in Experiments}
\label{sec:code}

In this appendix we present examples of the C code we have given to
the software model checkers in our verification experiments. The
reason we have used inequalities in the assertions is that the version
of Dagger we used did not support equalities.

Listing~\ref{modexpC} show the code corresponding to the
modular exponentiation algorithm shown in Figure~\ref{frenchmodexp},
where we have verified transcript-security modulo the hamming weight.

Listing~\ref{hammingManifest} presents the code used to show that the
hamming weight computation is a sound manifest declassifier.

\begin{lstlisting}[language=C,label=modexpC
                  ,caption={Verification code for modular exponentiation}]
#include <assert.h>

int modularExponentiation() {

  int m; // Length of the key
  int x1; // The secret 1
  int x2; // The secret 2

  int d1[m]; //Key 1
  int d2[m]; //Key 2

  //Hamming computation 1
  int hamming1 = 0;
  int i = m - 1;
  while(i >= 0) {
    hamming1 += (d1[i] ? 1 : 0);
    i--;
  }

  //Hamming computation 2
  int hamming2 = 0;
  int i = m - 1;
  while(i >= 0) {
    hamming2 += (d2[i] ? 1 : 0);
    i--;
  }

  //PROGRAM COPY 1
  int t1 = 0;
  int r1 = 1;
  int i = m - 1;
  int k = 0;
  while(i >= 0) {
    t1++;
    r1 = r1 * (k ? x1 : r1);
    k = k ^ d1[i];
    i = i - (k ? 0 : 1);
  }

  //PROGRAM COPY 2
  int t2 = 0;
  int r2 = 1;
  int i = m - 1;
  int k = 0;
  while(i >= 0) {
    t2++;
    r2 = r2 * (k ? x2 : r2);
    k = k ^ d2[i];
    i = i - (k ? 0 : 1);
  }

  if (hamming1 <= hamming2 && hamming1 >= hamming2) {
    assert(t1 <= t2);
    assert(t1 >= t2);
  }

  return r1;
}
\end{lstlisting}

\begin{lstlisting}[language=C,label=hammingManifest
                  ,caption={Verification code for the hamming weight declassifier}]
#include <assert.h>

int hammingManifest() {

  int m; //Length of the key

  int d1[m]; //Key 1
  int d2[m]; //Key 2

  //HAMMING COPY 1
  int i = m - 1;
  int hamming1 = 0;
  while(i >= 0) {
    hamming1 += (d1[i] ? 1 : 0);
    i--;
  }

  //HAMMING COPY 2
  int i = m - 1;
  int hamming2 = 0;
  while(i >= 0) {
    hamming2 += (d2[i] ? 1 : 0);
    i--;
  }

  //HAMMING^T COPY 1
  int i = m - 1;
  int hamming3 = 0;
  int t1 = 0;
  while(i >= 0) {
    hamming3 += (d1[i] ? 1 : 0);
    i--;
    t1++;
  }

  //HAMMING^T COPY 2
  int i = m - 1;
  int hamming4 = 0;
  int t2 = 0;
  while(i >= 0) {
    hamming4 += (d2[i] ? 1 : 0);
    i--;
    t2++;
  }

  if(hamming1 <= hamming2 && hamming1 >= hamming2) {
    assert(t1 <= t2);
    assert(t1 >= t2);
  }

  return 1;

}
\end{lstlisting}

\section{Another Example of Manifest Form}
\label{sec:manifestexample}

Here we present another example of an algorithm in manifest form
written in C. It is taken from \cite{Chevallier+:Low-cost}(figure 4a)
and is a two bit sliding window algorithm for modular exponentiation.

\begin{lstlisting}[language=C]
// m : Length of the key
// x : The secret
// d : The key, padded to length m+1

  // DECLASSIFIER
  int k = 1;
  int i = m;
  int s = 1;
  int l = 0;
  while(i >= 0) {
    k = (s ? 0 : 1) * (k+1);
    s = s ^ d[i+1] ^ (d[i] & (k % 2));
    i = i - k*s - (d[i] ? 0 : 1);
    l = l + 1;
  }


  //MAIN PROGRAM
  int R[3];

  R[0] = 1; R[1] = x; R[2] = x * x * x;
  d[0] = 0;
  int k = 1;
  int i = m;
  int s = 1;
  int j = l;

  while(j >= 0) {
    k = (!s) * (k+1);
    s = s ^ d[i+1] ^ (d[i] & (k % 2));
    R[0] = R[0] * R[k*s];

    i = i - k*s - (!d[i]);
    j = j - 1;
  }

  return R[0];
\end{lstlisting}

\fi

\end{document}